\documentclass[
 aip,
 jmp,
 amsmath,amssymb,
 reprint,
]{revtex4-2}

\usepackage{graphicx}
\usepackage{dcolumn}
\usepackage{bm}
\usepackage[mathlines]{lineno}

\begin{document}

\title[Synthetic Training and Representation Bridging in Reconstruction Domains]{Synthetic Training and Representation Bridging in Reconstruction Domains}

\author{Wonyong Chung}
\email{wonyongc@princeton.edu}
\affiliation{Department of Physics, Princeton University, New Jersey, USA}

\date{\today}

\begin{abstract}

Reconstructing low-dimensional truth labels from high-dimensional experimental data is a central challenge in any scenario that relies on robust mappings across this so-called domain gap, from multi-particle final states in high-energy physics to large-scale early-universe structure in cosmological surveys.
We introduce a new method to bridge this domain gap with an intermediate, synthetic representation of truth that differs from methods operating purely in latent space, such as normalizing flows or invertible approaches, in that the synthetic data is specifically engineered to represent intrinsic detector hardware capabilities of the system at hand.
The hypothesis is that by encoding physical properties of the detector response available only in full simulation, such synthetic representations result in a less lossy compression and recovery than a direct mapping from truth to experimental data.
We demonstrate a first implementation of this concept with full simulation of a dual-readout crystal electromagnetic calorimeter for future collider detectors, in which the synthetic data is constructed to be the simulated detector hits corresponding to photon tracks of scintillation and Cerenkov photons.
We refer to these signals as simulated observables as they would not be physical observables in a real detector, but are nonetheless representations of a real physical process.
First results show that the synthetic representation naturally anchors the neural network architecture to a known physical method, in this case the dual-readout correction.
We believe this strategy opens new avenues for machinistic interpretability and explainability of ML-based reconstruction methods. In the case of anomalous signal detection, we hypothesize that anomalous signals detected in networks trained on synthetic data rooted in a physical process are more likely to be indicative of a genuinely physical anomaly.

\end{abstract}

\maketitle

\section{Introduction}
High-energy physics detectors produce vast, high-dimensional data from particle collisions.
In calorimeters alone, each event can generate hundreds to thousands of hit signals, encoding energy depositions in intricate spatial patterns.
Such complexity contrasts sharply with the relatively low-dimensional space of Monte Carlo (MC) truth labels—typically consisting of a particle type, momentum, and vertex—used for training reconstruction and classification algorithms.
This disparity in dimensionality often leads to degenerate mappings: multiple calorimeter hit patterns can correspond to the same nominal MC truth label.

While modern machine learning techniques excel at extracting robust features from high-dimensional data, their performance can still be limited by the fundamental mismatch between compact, idealized truth and the rich detector response.
Methods such as fast simulation with generative adversarial networks (GANs) or normalizing flows have attempted to learn transformations from particle-level truth to realistic calorimeter data, with varying degrees of success and fidelity to real detector effects~\cite{atlas_deepGAN,calogan,gan4hls}.
Likewise, invertible ML approaches (e.g., OmniFold)~\cite{omnifold,ml_landscape} aim to invert the detector response to recover underlying truth-level information.
Yet many of these frameworks implicitly treat MC truth as a low-dimensional label, leaving significant degeneracies when distinguishing topologically similar shower patterns.

We propose a new method to address the dimensionality gap and associated information loss.
Instead of directly learning the mapping between the low-dimensional truth space $\mathbf{T}$ and the high-dimensional detector space $\mathbf{D}$, we introduce a synthetic, intermediate representation of truth, denoted space $\mathbf{S}$.
This intermediate space is designed to be "closer" to the detector response space $\mathbf{D}$ in terms of dimensionality and information content.
Crucially, elements of $\mathbf{S}$ are constructed based on detector-specific hardware capabilities and are therefore representations of a real physical process that are visible only in full simulation.
The core hypothesis is that by factorizing the inference problem into two stages, $\mathbf{D}\rightarrow\mathbf{S}\rightarrow\mathbf{T}$, the overall process becomes less susceptible to information loss.
The mapping between the two high-dimensional spaces $\mathbf{D}\rightarrow\mathbf{S}$ is expected to be less degenerate and more information-preserving than the direct mapping $\mathbf{D}\rightarrow\mathbf{T}$.
The subsequent mapping $\mathbf{S}\rightarrow\mathbf{T}$ then operates on a richer, potentially better-conditioned feature space.

In this work, elements of $\mathbf{S}$ are instantiated as detector cell hits corresponding to the trajectories of optical Cerenkov (C) and scintillation (S) photons within longitudinally segmented dual-readout calorimeter crystals. These would not be real observables in a real detector, but here we refer to them as simulated observables, as they nonetheless represent a true physical process.
This approach encodes the intrinsic physics principles of dual-readout calorimetry as well as the segmented geometry of the detector into the space $\mathbf{S}$. The intent is to surface the hidden features of shower development in calorimetry which underpin reconstruction algorithms but are not directly observable, such skin-depth and fine-grained shower structure.

\begin{figure}[h]
    \centering
    \includegraphics[width=0.6\textwidth]{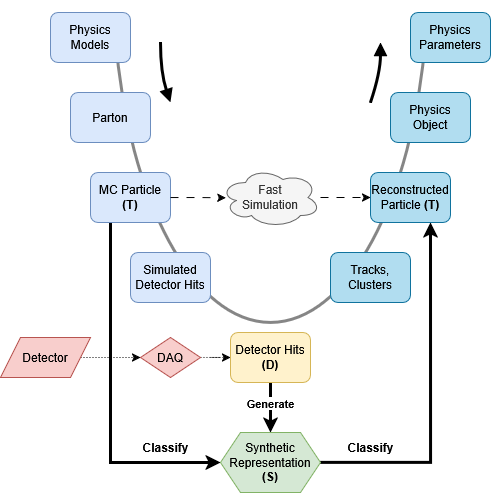}
    \caption{The U-Net structure of the traditional detector simulation chain is shown with the conceptual extension of the synthetic space $\mathbf{S}$. Starting from physics models, event generators and hadronization programs produce the MC particles ($\mathbf{T}$) to be simulated in a detector. Simulated detector hits are digitized to the space of realistic detector hits $\mathbf{D}$ produced by a real detector and DAQ. Raw detector hits typically undergo a process of track forming and clusterization for particle-flow algorithms to process them into reconstructed particles ($\mathbf{T}$) from which physics objects and parameters are extracted. The U-Net name comes from the process of folding dimensionality from one layer to the next, up to a so-called bottleneck layer (in this case, the detector hits $\mathbf{D}$), an arbitrary point at which global context is aggregated and skip-connections reconstruct features at matching scales on the way back up to the original domain (or equivalent). The synthetic representations $\mathbf{S}$ sit at a level below the detector hits and are engineered to encode additional information about the detector response available only in full simulation. The core hypothesis is that by basing the synthetic representation on intrinsic detector capabilities and known physics processes, the mapping $\mathbf{S}\rightarrow\mathbf{T}$, while still bridging a large dimensional-gap, is better-conditioned and less lossy than the direct connection $\mathbf{D}\rightarrow\mathbf{T}$.}
    \label{detector_simulation_chain}
\end{figure}

The primary goal of this initial work is to establish that a machine learning model can successfully learn the mapping from realistic, observable detector hits ($\mathbf{D}$, comprising energy deposits above threshold and S/C photon counts) to the synthetic, intermediate representation based on S/C optical photon tracks. A U-Net architecture is employed for this image-to-image translation task. Qualitative results are presented, suggesting that the learned mapping preserves and reconstructs not only fine-grained topological information present in the synthetic representation, but does so according to the principles of dual-readout correction that underlie the construction of the synthetic representation. We interpret these results as being significant for machinistic interpretability of reconstruction algorithms and their physical context.

\section{Methodology}

Denote the three spaces as follows:

\begin{itemize}
    \item \( \mathbf{T} \in \mathbb{R}^{d} \): the traditional MC truth label space (e.g., particle type, momentum, vertex).
    \item \( \mathbf{S} \in \mathbb{R}^{m} \): the synthetic intermediate representation space, generated by a hardware-specific detector response function
    \item \( \mathbf{D} \in \mathbb{R}^{n} \): the space of realistic detector hits likely to be observed in a real experiment, with \( n \gg d \) and typically \( m \gtrsim n \) or at least \( m \gg d \).
\end{itemize}
 
We train two neural networks:
 
\begin{itemize}
    \item \textbf{NN1:}  \( f: \mathbf{D} \to \mathbf{S} \). This network is a forward mapping trained on pairs \( (D_i, S_i) \) to predict a synthetic intermediate representation from a given set of realistic hits. Architecturally, this can be implemented as an encoder-decoder model or a pair of generative/discriminative networks (e.g., using adversarial objectives).
    \item \textbf{NN2:}  \( g: \mathbf{S} \leftrightarrow \mathbf{T} \). This network is a forward-backward mapping trained on pairs \( (S_i, T_i) \) to perform classification and/or regression to learn and recover a traditional MC truth from a synthetic representation.
\end{itemize}

Given a new set of detector hits $D_j$, $\mathrm{NN}_1$ infers its synthetic representation $S_j$. $\mathrm{NN}_2$ then classifies $S_j$ to yield the final label $T_j$.
By factoring the mapping $\mathbf{D}\to \mathbf{T}$ through $\mathbf{S}$, we aim to anchor the convolutions in latent space to a detector-specific physical process.
Formally, given a new detector hit collection $D_1$, we compute
$$
 S_1 = f(D_1) \quad \text{and then} \quad T_1 = g(S_1), 
$$

so that the composite mapping $g\circ f: \mathbf{D} \to \mathbf{T}$ is used for particle identification or other reconstruction tasks.
 
Let $L_T$ and $L_S$ be appropriate loss functions (e.g., cross-entropy for classification, mean-squared error for regression). The two training objectives are:
$$
 \mathcal{L}_{S} = \mathbb{E}_{(D,S)}\left[ L_S\bigl(f(D),\,S\bigr) \right],
$$
$$
 \mathcal{L}_{T} = \mathbb{E}_{(S,T)}\left[ L_T\bigl(g(S),\,T\bigr) \right]. 
$$
The overall loss (when considering the full pipeline) is
$$
 \mathcal{L}_{\text{total}} = \mathbb{E}_{(D,S,T)}\left[ L_T\bigl(g(f(D)),\,T\bigr) \right] + \lambda\,\mathcal{L}_{S}, 
$$
where $\lambda$ is a hyperparameter balancing the two losses. Although we presently train the networks in two stages, an end-to-end training scheme is also conceivable.

\section{Detector Description and Differentiable Full Simulation}
The simulations in this study were performed using a custom full detector simulation implemented in the key4hep software ecosystem for future collider detectors.
The detector geometry and response functions are implemented in DD4hep, a modularized wrapper around the Geant4 simulation package for particle interactions with matter.
A custom edm4hep data class is used for the synthetic data.
This framework provides a flexible and comprehensive environment for simulating detector concepts proposed for future collider experiments.

\begin{figure}[h]
    \centering
    \includegraphics[width=\textwidth]{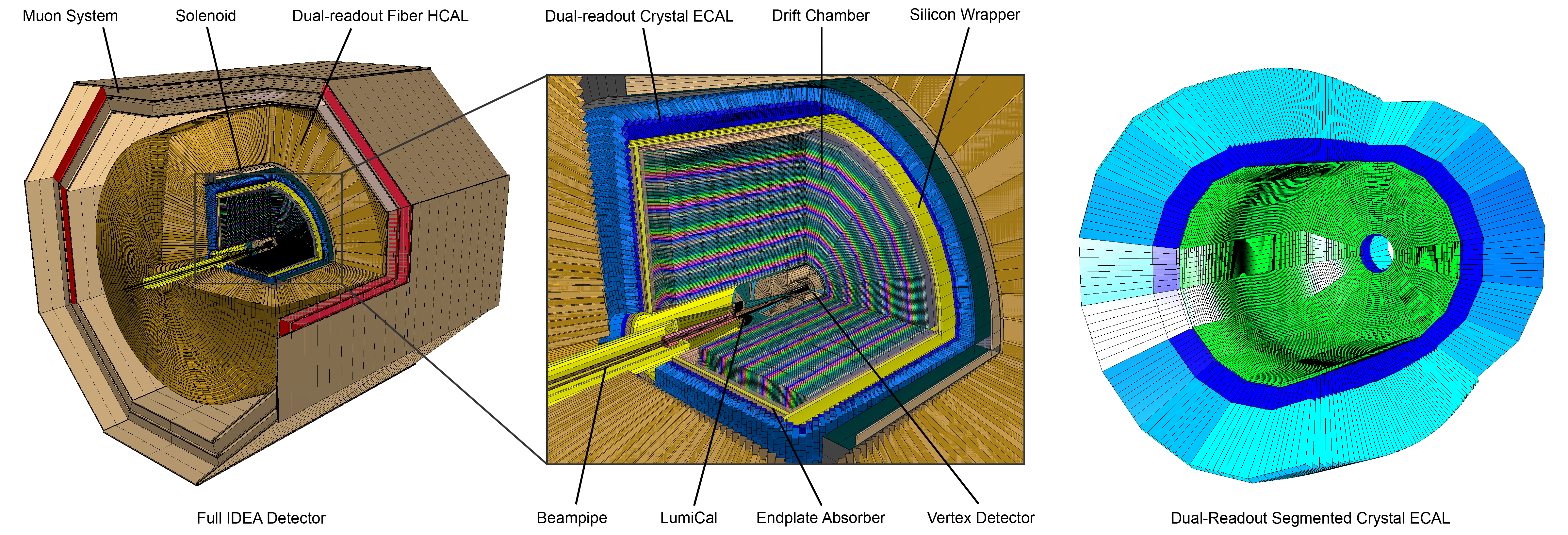}
    \caption{Left and middle panes: The full IDEA detector showing all subdetectors with labels. Right pane: isolated view of only the dual-readout segmented crystal electromagnetic calorimeter used for this study. Crystal dimensions are enlarged by a factor of 10x for visibility.}
    \label{IDEA_SCEPCal}
\end{figure}

\begin{figure}[h]
    \centering
    \includegraphics[width=\textwidth]{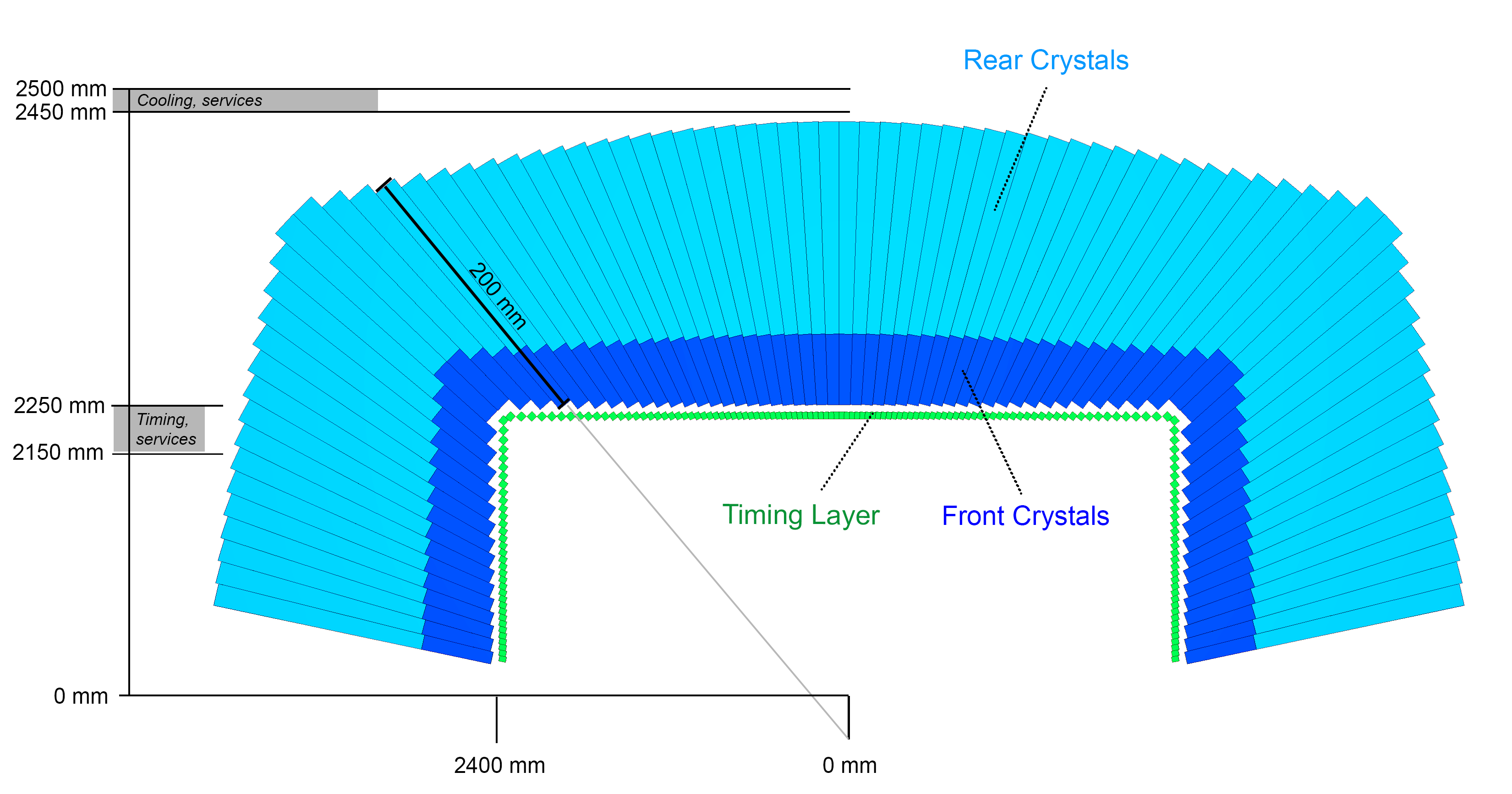}
    \caption{A single phi-slice of the dual-readout segmented crystal calorimeter used for this study. The main crystal layer (comprising front and rear crystals) is instrumented with a projective offset to point away from the interaction point as shown, to mitigate particle leakage through projective cracks. To maximize angular resolution, the timing layer has no projective offset. The timing layer is not used for this study.}
    \label{scepcal_phislice}
\end{figure}

The detector is a dual-readout, homogeneous crystal electromagnetic calorimeter set in a projective geometry. A single phi-slice of the detector is shown in Figure~\ref{scepcal_phislice}.
The crystals are longitudinally segmented into separate front and rear crystals with separate readout channels for scintillation and Cerenkov light. 
The nominal transverse granularity of the crystals is 1x1\,cm, with the front and rear crystal lengths in a ratio of 6:16 radiation lengths in depth for a total of 22 radiation lengths for full EM shower containment.
The crystal material for this study is lead tungstate, used in the CMS experiment for its short radiation length and moderate scintillation yield.
At a future $e^+e^-$ collider, this ECAL is designed to be paired with the IDEA dual-readout fiber calorimeter acting as the hadronic calorimeter (HCAL) and is part of the new IDEA baseline design shown in Figure~\ref{IDEA_SCEPCal}.
When paired with the IDEA HCAL, the ECAL was shown in a previous work~\cite{Lucchini_2020} to significantly improve hadronic jet resolution, alongside providing excellent intrinsic EM resolution.
A precision timing layer providing picosecond time resolution is also envisioned to be instrumented in front of the main crystal layer, assumed to be made of a fast scintillating crystal such as LYSO, although LYSO is more suited for a proton-proton collider environment such as at the LHC.
The timing layer is not used for this study and the choice of material is under study.
The main projective layer mitigates projective cracks with a radial pointing offset around the interaction point similar to the CMS detector. The timing layer does not use a projective offset to maximize angular resolution.

The detector geometry is written to be dynamically reconfigurable for use in a differentiable full simulation pipeline, meaning the detector construction has no hard-coded parameters and is fully parameterized to a few input parameters such as the overall dimensions and nominal crystal granularities, including the option to bundle different numbers of front and rear crystals together in a single crystal tower.
The individual dimensions and projective placements of the crystals are then automatically calculated.
The number of front to rear crystals per tower is 1:1 in this study.

To our knowledge, this simulation is the first fully dynamic and reconfigurable geometry written for a collider detector and enables systematic studies of detector geometry optimizations alongside reconstruction algorith parameters in a process called bilevel optimization.
More details of the detector construction can be found in~\cite{calor2024}.

Optical surface definitions are implemented on the crystals and amount to defining the reflection and transmission coefficients of the material.
However, the effect on the simulated outputs was found to be minimal at the expense of a significant increase in compute, so for this study the optical surface effects were turned off.

The standard output of a DD4hep simulation is edm4hep calorimeter hits, which store accumulated energy deposit per event in each detector cell, or crystal.
For dual-readout calorimetery, we are interested in optical photons, so we additionally implement a custom readout class to save the number of scintillation and Cerenkov photons produced per event in each detector cell, representing the realistic detector response in the spacee $D$.

\section{Synthetic Representation of Detector Response}

An important capability of full simulation which has traditionally been overlooked is the possibility to customize the detector response function, which runs at every step of every particle in the simulation and contains the logic of when to create and save a detector hit.
The typical approach is to save hits based on the energy deposited in the material at each step of the simulation, subject to a threshold value, usually 1\,keV, reflecting the finite sensitivity and noise thresholds of real sensors.

In addition to the counts of S/C photons produced per crystal, we implement a custom detector response function to generate a synthetic detector response in the space $S$.

Instead of applying an energy-deposit cut at the step-level, we apply a cut at the track-level for optical photons based on the wavelength of the photon, registering a detector hit for photons with a wavelength within the specified range, even if the energy deposited in a particular step is zero, effectively saving the tracks of optical photons through the detector as blank, zero-energy hits.
In this study we use the wavelength range 200-600\,nm in accordance with the scintillation and cerenkov production wavelengths of lead tungstate.

\begin{figure}[h]
    \centering
    \includegraphics[width=\textwidth]{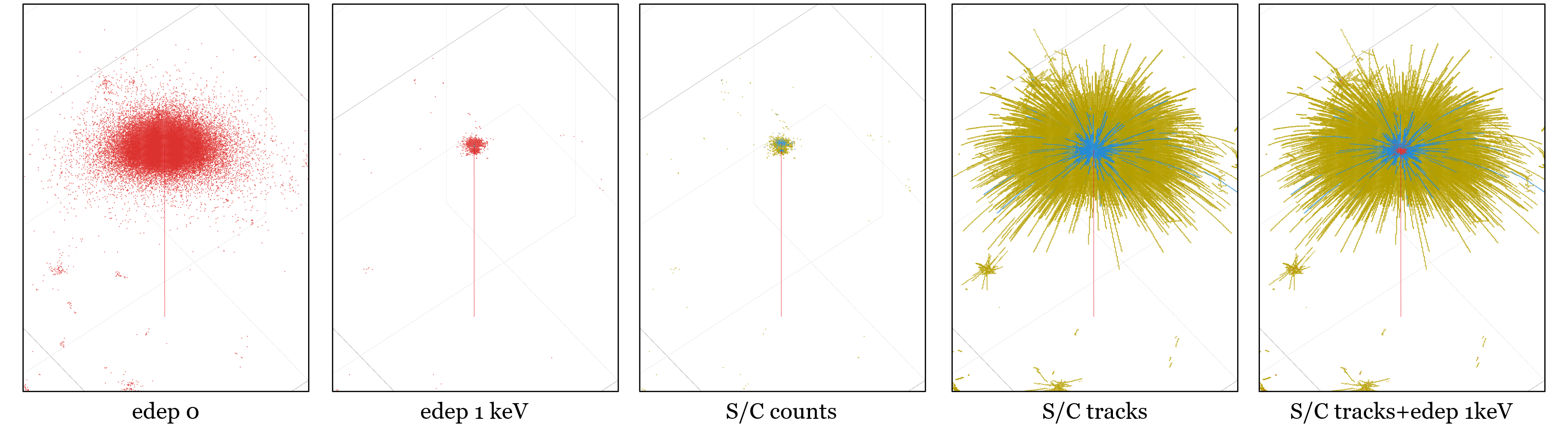}
    \caption{Event displays for a single 50\,GeV electron showing detector hits for realistic and synthetic detector response functions. Each point is a single crystal hit. From left to right: edep0 results in a very dense cloud of hits as every step with non-zero energy deposit is saved. Using edep 1\,keV or non-zero S/C photon counts results in a much more concentrated cloud of hits and is most representative of a realistic detector response. Next, saving all hits belonging to a S/C photon track, even if the energy deposit is zero, produces an image with rich geometrical structure encoding properties of the detector interaction visible only in full simulation. Finally, superimposing edep 1\,keV on top of S/C tracks illustrates the dimensional-gap we propose to bridge.}
    \label{detector_responses}
\end{figure}

The event displays in Figure~\ref{detector_responses} illustrate the core idea of this work, which is that the synthetic representations offer a rich source of structural information which are unphysical in the sense that one would not ever see them in a real detector, but they are not entirely unphysical in that they are representations of a true physical process tied to the physical capabilities of the detector hardware, in this case, dual-readout and longitudinal segmentation. 

The hypothesis is that because this synthetic representation is grounded in the detector hardware capabilities rather than existing solely in abstract latent space, it can act as a less lossy bridge between the realistic detector hits $D$ and the truth labels $T$. The mapping $T\rightarrow S$ is effectively a transformation of the labels into the space of pure detector response, where particles are no longer described by a particle type and momentum, but rather as rich geometrical structures of detector hits in a synthetic space only available in full simulation.

\section{Event Generation and Image Encoding}

Single particle events are run using the built-in DD4hep/Geant4 particle gun. We simulate 10,000 events each of electrons, photons (gammas), $\pi^0$, $\pi^+$, $\pi^-$, and neutrons, all emitted at a fixed momentum of 50 GeV from the interaction point with no smearing, in a uniform angular distribution in the barrel of the detector only. We determine 10,000 events to be a sufficient sample size for this study because single-particle events have a high degree of self-similarity and the intent is to first establish a handle in hyperparameter space and perform qualitative examinations of the method.

\begin{figure}[h]
    \centering
    \includegraphics[width=\textwidth]{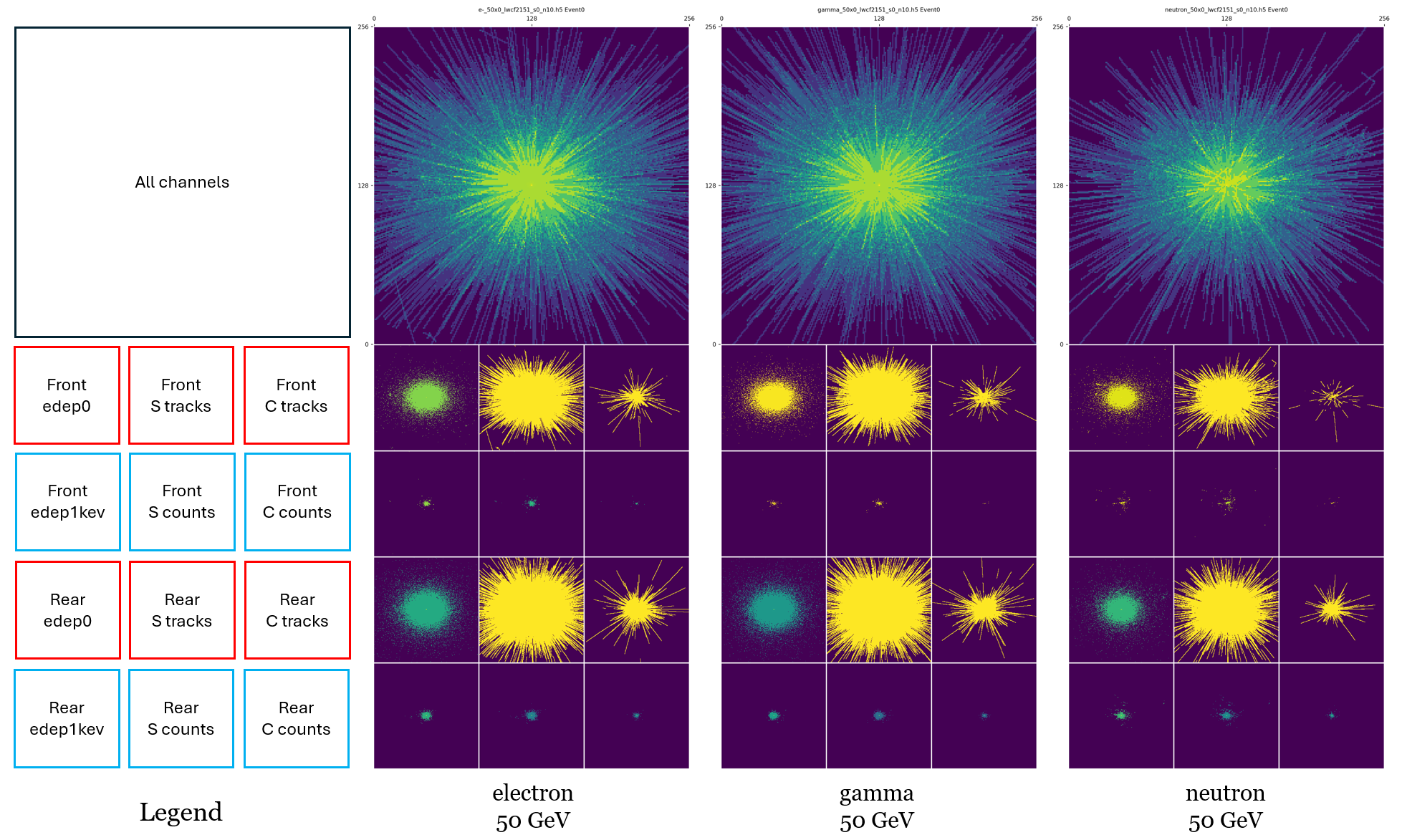}
    \caption{Representative samples of constructed images for 50\,GeV electrons, gammas, and neutrons. The sum of all channels per pixel is displayed in the top pane for each particle type, and the 12 channels shown separately below, corresponding to the legend at left. The channels boxed in red in the legend are those considered synthetic, and blue channels represent realistic detector responses. Compared to the realistic channels, the synthetic channels, S/C tracks in particular, offer a rich geometrical structure which encode aspects of detector response visible only in full simulation.}
    \label{image_gen_egn}
\end{figure}

\begin{figure}[h]
    \centering
    \includegraphics[width=\textwidth]{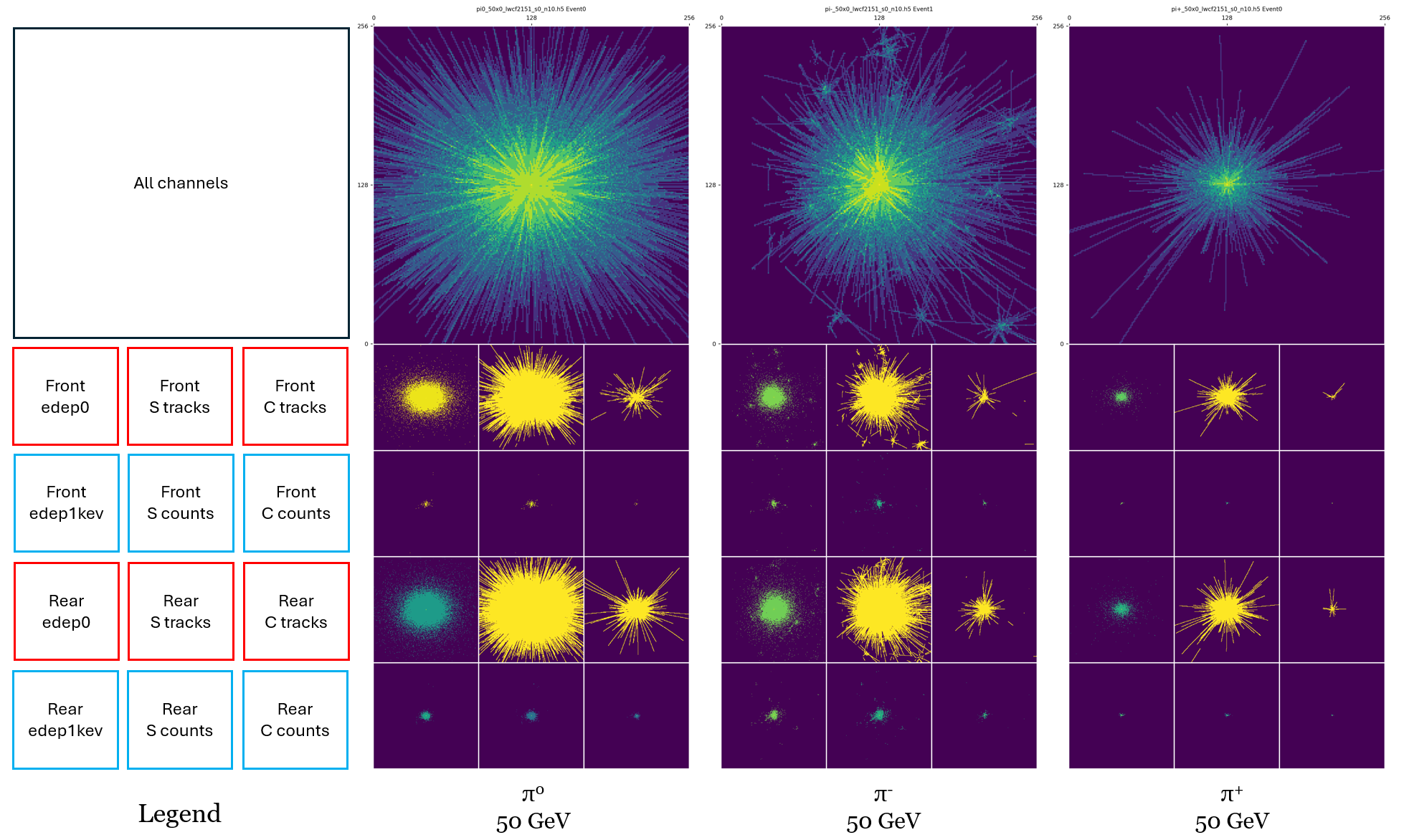}
    \caption{Representative samples of constructed images for 50\,GeV neutral and charged pions. Neutral pions typically decay almost immediately into two photons, and resolving the signature of two overlapping photons underpins $\pi^0$/gamma separation. In contrast, charged pions may shower early or late in the ECAL. In these samples, the $\pi^-$, middle, exhibits a typical pattern of early-showing, with multiple secondaries around the main island clearly visible. The $\pi^+$, right, is a typical late-shower, with a noticeably sparse scintillation signal.}
    \label{image_gen_pions}
\end{figure}

A 12-channel image is constructed for each event using the following procedure.
First, the hit with the maximum energy deposit in the event is identified and a breadth-first-search starting from this hit is conducted to collect the set of directly neighboring detector hits, including depth, comprising the "island" of hits originating from the maximum-energy hit.
A fixed-size 256\,x\,256 pixel image is then constructed for the island of hits, centered around the maximum-energy hit, with each pixel representing a single crystal tower of one front and one rear crystal. Each pixel is encoded with 12 channels of data, 6 for the front crystal and 6 for the rear crystal. The channel descriptions and values are listed in Table~\ref{tab:image_channels}. Representative samples of images constructed for electrons, gammas, and neutrons are shown in Figure~\ref{image_gen_egn}, and neutral and charged pions in Figure~\ref{image_gen_pions}.

\begin{table}
\caption{\label{tab:image_channels}Channel encoding scheme for each event image. $E$, $S$, and $C$ are the sum of the energy deposits and S/C photon counts for the island of hits. $e_i^0$, $e_i^{\text{1 keV}}$, $s_i$, and $c_i$ are the energy deposits (using edep0 and edep1keV) and S/C photon counts for the single crystal at each pixel. A log-weighted formula, which compresses small signals, is used to normalize the dynamic range of the energy deposits and S/C counts, with $s_e$, $s_p$, $w_e$, $w_p$ as scaling and offset factors. In this study $s_e=2$, $s_p=5$, and $w_e=w_p=1$. An alternative weighting scheme such as a square-root function or arcsine may be used to preserve signal linearity, however to first-order, the choice of weighting function is not expected to be significant for a qualitative assessment of the method.}

\begin{ruledtabular}
\begin{tabular}{lll}
Channel & Description & Value \\ 
    \hline 
    1  & Front crystal edep0 & $s_e\,\log(e_i^0/E+1)+w_e$ \\
    2  & Front S track hits  & 1 if present, 0 otherwise \\
    3  & Front C track hits  & 1 if present, 0 otherwise \\
    4  & Front edep1 keV     & $s_e\,\log(e_i^{\text{1 keV}}/E+1)+w_e$ \\
    5  & Front S counts      & $s_p\,\log(s_i/S+1)+w_p$ \\
    6  & Front C counts      & $s_p\,\log(c_i/C+1)+w_p$ \\
    7  & Rear crystal edep0  & $s_e\,\log(e_i^0/E+1)+w_e$ \\
    8  & Rear S track hits   & 1 if present, 0 otherwise \\
    9  & Rear C track hits   & 1 if present, 0 otherwise \\
    10 & Rear edep1 keV      & $s_e\,\log(e_i^{\text{1 keV}}/E+1)+w_e$ \\
    11 & Rear S counts       & $s_p\,\log(s_i/S+1)+w_p$ \\
    12 & Rear C counts       & $s_p\,\log(c_i/C+1)+w_p$ \\
    \hline 
\end{tabular}
\end{ruledtabular}
\end{table}

\section{First Implementation: A 3-level U-Net}

For a first test of this idea we choose a U-Net architecture to implement the image-to-image translation $D\rightarrow S$, extending the traditional structure of the detector simulation chain as shown in Figure~\ref{detector_simulation_chain}.

We implement a 3-level U-Net to train over the 10,000 images for each particle in various combinations (just electrons, or electrons and gammas, etc.). The training loop makes 2 copies of each image: full and masked. The masked image zeros-out the synthetic channels and the model is trained to infer the missing synthetic channels when given an image with only the realistic channels present. The loss function used is a weighted L1+SSIM loss, which, concisely, considers a combination of absolute pixel difference (L1) and structural correlations (SSIM) within the image. We train over a modest hyperparameter space with batch sizes 2, 4, 8, or 16, with an initial learning rate of $1e-3$ with a scheduler halving the rate to a minimum of $1e-7$ with a patience of 5. The training loop is run for 500 epochs. Then we run inference to generate images for a test set of 1000 images for each particle type and classify them.

\section{Inference: First Look}
\begin{figure}[h]
    \centering
    \includegraphics[width=\textwidth]{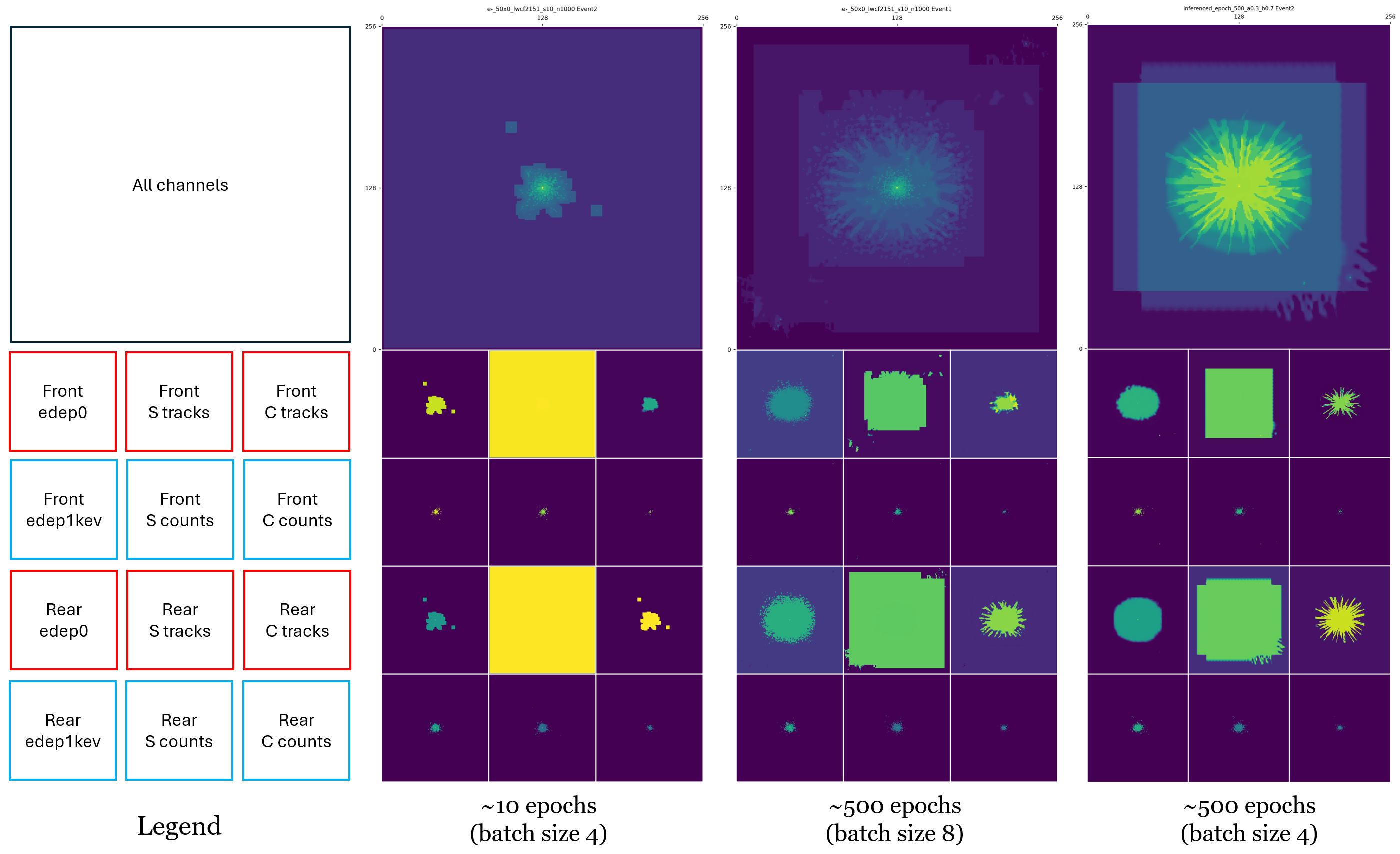}
    \caption{Inference images of a 50\,GeV electron after various training epochs and batch sizes. See text.}
    \label{inference_firstlook}
\end{figure}

A sample of inferenced images for various hyperparameters are shown in Figure~\ref{inference_firstlook}.
While tuning hyperparameters is clearly the challenge for a production-grade model, here we discuss qualitative assesments of the model's progression and interpret its output.

The performance of the model appears to depend most significantly on the batch size, i.e. the number of simultaneous images over which the model attempts to calculate a gradient and apply the loss function.
As noted previously, the constructed images have an inherently high degree of self-similarly because they effectively encode the stochastic differences in detector response from the same type of single-particle event.
The small structural differences we are interested in discriminating are likely to be washed out if a large batch size is used. Conversely, using the minimum batch size of 2 would in principle be most effective at discerning fine, pixel-level differences, however the tradeoff here is that the model may become stuck in the local extrema of these features and never converge to the global context.
Using a modest batch size of 2-4 with a more robust scheduler is likely to be the most direct path to improvement.

Nonetheless, we observe that the cerenkov signal begins to be resolved first across all variations, a robust indication of the model's explainability.
In the context of image-to-image translation, this is also an expected result because the Cerenkov signal is in general much more sparse than the scintillation signal and therefore, assuming all channels of the image are weighted equally in the loss, one would expect the model to pick up on the gradient landscape of the Cerenkov signal during feedforward and backpropagation more quickly than the scintillation signal.

Conversely, the scintillation signals already nearly saturate the available canvas before being zeroed-out, and subsequently, in early epochs, appear as a large uniform square, reflecting the difficulty of feature resolution in this regime.
In later epochs we observe structure beginning to emerge in the scintillation signal from the outside-in, indicating that the radial structure of the tracks, in particular outlier "hair"-like protrusions, offers the most discriminating power.
Given the negative effects of a saturated signal, we note that it may be advantageous to attentuate the scintillation signal for this type of analysis, contrary to the traditional wisdom in non-dual-readout systems that that more signal is always better. 

Our interpretation of these results is that the model is effectively machine-learning the dual-readout correction factor of the detector.
In a more traditional analysis, one would calibrate the detector on known electromagnetic and hadronic processes separately to obtain the respective S/C scaling response factors, then use these to determine the EM fraction of the shower, mitigating the event-by-event hadronic shower fluctuations seen in non-dual-readout calorimeters.
By locking on to the sparse Cerenkov signal first while the more degenerate scintillation signal is resolved slowly, we interpret the model to be implicitly performing the dual-readout correction.
Because we have deliberately engineered the synthetic data to represent the known physical process of dual-readout rooted in the specific detector geometry, we show for the first time that a simulated observables can augment real observables in an interpretable and explainable way. We posit that in systems engineering in such a fashion, anomalous ML signals may be more likely to genuinely represent physically anomalous signals.

\section{Outlook}
We have introduced a new method of encoding simulated observables into a space of synthetic representations based on intrinsic hardware capabilities and real physical processes.
The concept is demonstrated to be interpretable in machine learning models with respect to an established instrumentation technique, in this study, dual-readout calorimetry.
We expect the core idea introduced here to be applicable to a wide range of experiments that rely on high-statistics Monte Carlo simulations mapping low-dimensional truth-labels to high-dimensional, complex detector responses.

For the image-to-image translation task of this specific work we have used a U-Net model as a natural and easy conceptual extension of the detector simulation chain.
In general, any suitable architecture may be used as long as the motivations underlying the synthetic response are grounded in real physical processes and detector capabilities.
For image generation, denoising latent diffusion models are an attractive next avenue for testing and offer distinct advantages with respect to the possibility of encoding detector systematics directly into the model.
Although timing information was not used in this study, vector sequencing methods such as GNNs or transformers may be used in a similar way in studies using timing information or other dimensional information specific to the use case.

\begin{acknowledgments}
This work was supported in part by the U.S. DOE Office of Science, Office of High Energy Physics under Contract No. DE-SC0022045, Maximal Information Calorimetry.
\end{acknowledgments}

\bibliography{main}

\begin{thebibliography}{7}%
\makeatletter
\providecommand \@ifxundefined [1]{%
 \@ifx{#1\undefined}
}%
\providecommand \@ifnum [1]{%
 \ifnum #1\expandafter \@firstoftwo
 \else \expandafter \@secondoftwo
 \fi
}%
\providecommand \@ifx [1]{%
 \ifx #1\expandafter \@firstoftwo
 \else \expandafter \@secondoftwo
 \fi
}%
\providecommand \natexlab [1]{#1}%
\providecommand \enquote  [1]{``#1''}%
\providecommand \bibnamefont  [1]{#1}%
\providecommand \bibfnamefont [1]{#1}%
\providecommand \citenamefont [1]{#1}%
\providecommand \href@noop [0]{\@secondoftwo}%
\providecommand \href [0]{\begingroup \@sanitize@url \@href}%
\providecommand \@href[1]{\@@startlink{#1}\@@href}%
\providecommand \@@href[1]{\endgroup#1\@@endlink}%
\providecommand \@sanitize@url [0]{\catcode `\\12\catcode `\$12\catcode `\&12\catcode `\#12\catcode `\^12\catcode `\_12\catcode `\%12\relax}%
\providecommand \@@startlink[1]{}%
\providecommand \@@endlink[0]{}%
\providecommand \url  [0]{\begingroup\@sanitize@url \@url }%
\providecommand \@url [1]{\endgroup\@href {#1}{\urlprefix }}%
\providecommand \urlprefix  [0]{URL }%
\providecommand \Eprint [0]{\href }%
\providecommand \doibase [0]{https://doi.org/}%
\providecommand \selectlanguage [0]{\@gobble}%
\providecommand \bibinfo  [0]{\@secondoftwo}%
\providecommand \bibfield  [0]{\@secondoftwo}%
\providecommand \translation [1]{[#1]}%
\providecommand \BibitemOpen [0]{}%
\providecommand \bibitemStop [0]{}%
\providecommand \bibitemNoStop [0]{.\EOS\space}%
\providecommand \EOS [0]{\spacefactor3000\relax}%
\providecommand \BibitemShut  [1]{\csname bibitem#1\endcsname}%
\let\auto@bib@innerbib\@empty
\bibitem [{atl(2024)}]{atlas_deepGAN}%
  \BibitemOpen
  \bibfield  {title} {\enquote {\bibinfo {title} {{Deep Generative Models for Fast Photon Shower Simulation in ATLAS}},}\ }\href {https://doi.org/10.1007/s41781-023-00106-9} {\bibfield  {journal} {\bibinfo  {journal} {Comput. Softw. Big Sci.}\ }\textbf {\bibinfo {volume} {8}},\ \bibinfo {pages} {7} (\bibinfo {year} {2024})},\ \Eprint {https://arxiv.org/abs/2210.06204} {arXiv:2210.06204} \BibitemShut {NoStop}%
\bibitem [{\citenamefont {Paganini}, \citenamefont {de~Oliveira},\ and\ \citenamefont {Nachman}(2018)}]{calogan}%
  \BibitemOpen
  \bibfield  {author} {\bibinfo {author} {\bibfnamefont {M.}~\bibnamefont {Paganini}}, \bibinfo {author} {\bibfnamefont {L.}~\bibnamefont {de~Oliveira}},\ and\ \bibinfo {author} {\bibfnamefont {B.}~\bibnamefont {Nachman}},\ }\bibfield  {title} {\enquote {\bibinfo {title} {Calogan: Simulating 3d high energy particle showers in multilayer electromagnetic calorimeters with generative adversarial networks},}\ }\href {https://doi.org/10.1103/PhysRevD.97.014021} {\bibfield  {journal} {\bibinfo  {journal} {Phys. Rev. D}\ }\textbf {\bibinfo {volume} {97}},\ \bibinfo {pages} {014021} (\bibinfo {year} {2018})}\BibitemShut {NoStop}%
\bibitem [{\citenamefont {Musella}\ and\ \citenamefont {Pandolfi}(2018)}]{gan4hls}%
  \BibitemOpen
  \bibfield  {author} {\bibinfo {author} {\bibfnamefont {P.}~\bibnamefont {Musella}}\ and\ \bibinfo {author} {\bibfnamefont {F.}~\bibnamefont {Pandolfi}},\ }\bibfield  {title} {\enquote {\bibinfo {title} {{Fast and Accurate Simulation of Particle Detectors Using Generative Adversarial Networks}},}\ }\href {https://doi.org/10.1007/s41781-018-0015-y} {\bibfield  {journal} {\bibinfo  {journal} {Comput. Softw. Big Sci.}\ }\textbf {\bibinfo {volume} {2}},\ \bibinfo {pages} {8} (\bibinfo {year} {2018})},\ \Eprint {https://arxiv.org/abs/1805.00850} {arXiv:1805.00850 [hep-ex]} \BibitemShut {NoStop}%
\bibitem [{\citenamefont {Andreassen}\ \emph {et~al.}(2020)\citenamefont {Andreassen}, \citenamefont {Komiske}, \citenamefont {Metodiev}, \citenamefont {Nachman},\ and\ \citenamefont {Thaler}}]{omnifold}%
  \BibitemOpen
  \bibfield  {author} {\bibinfo {author} {\bibfnamefont {A.}~\bibnamefont {Andreassen}}, \bibinfo {author} {\bibfnamefont {P.~T.}\ \bibnamefont {Komiske}}, \bibinfo {author} {\bibfnamefont {E.~M.}\ \bibnamefont {Metodiev}}, \bibinfo {author} {\bibfnamefont {B.}~\bibnamefont {Nachman}},\ and\ \bibinfo {author} {\bibfnamefont {J.}~\bibnamefont {Thaler}},\ }\bibfield  {title} {\enquote {\bibinfo {title} {Omnifold: A method to simultaneously unfold all observables},}\ }\href {https://doi.org/10.1103/PhysRevLett.124.182001} {\bibfield  {journal} {\bibinfo  {journal} {Phys. Rev. Lett.}\ }\textbf {\bibinfo {volume} {124}},\ \bibinfo {pages} {182001} (\bibinfo {year} {2020})}\BibitemShut {NoStop}%
\bibitem [{\citenamefont {Huetsch}\ \emph {et~al.}(2025)\citenamefont {Huetsch}, \citenamefont {Villadamigo}, \citenamefont {Shmakov}, \citenamefont {Diefenbacher}, \citenamefont {Mikuni}, \citenamefont {Heimel}, \citenamefont {Fenton}, \citenamefont {Greif}, \citenamefont {Nachman}, \citenamefont {Whiteson}, \citenamefont {Butter},\ and\ \citenamefont {Plehn}}]{ml_landscape}%
  \BibitemOpen
  \bibfield  {author} {\bibinfo {author} {\bibfnamefont {N.}~\bibnamefont {Huetsch}}, \bibinfo {author} {\bibfnamefont {J.~M.}\ \bibnamefont {Villadamigo}}, \bibinfo {author} {\bibfnamefont {A.}~\bibnamefont {Shmakov}}, \bibinfo {author} {\bibfnamefont {S.}~\bibnamefont {Diefenbacher}}, \bibinfo {author} {\bibfnamefont {V.}~\bibnamefont {Mikuni}}, \bibinfo {author} {\bibfnamefont {T.}~\bibnamefont {Heimel}}, \bibinfo {author} {\bibfnamefont {M.}~\bibnamefont {Fenton}}, \bibinfo {author} {\bibfnamefont {K.}~\bibnamefont {Greif}}, \bibinfo {author} {\bibfnamefont {B.}~\bibnamefont {Nachman}}, \bibinfo {author} {\bibfnamefont {D.}~\bibnamefont {Whiteson}}, \bibinfo {author} {\bibfnamefont {A.}~\bibnamefont {Butter}},\ and\ \bibinfo {author} {\bibfnamefont {T.}~\bibnamefont {Plehn}},\ }\bibfield  {title} {\enquote {\bibinfo {title} {{The landscape of unfolding with machine learning}},}\ }\href {https://doi.org/10.21468/SciPostPhys.18.2.070} {\bibfield  {journal} {\bibinfo  {journal} {SciPost Phys.}\ }\textbf {\bibinfo {volume} {18}},\ \bibinfo {pages} {070} (\bibinfo {year} {2025})}\BibitemShut {NoStop}%
\bibitem [{\citenamefont {Lucchini}\ \emph {et~al.}(2020)\citenamefont {Lucchini}, \citenamefont {Chung}, \citenamefont {Eno}, \citenamefont {Lai}, \citenamefont {Lucchini}, \citenamefont {Nguyen},\ and\ \citenamefont {Tully}}]{Lucchini_2020}%
  \BibitemOpen
  \bibfield  {author} {\bibinfo {author} {\bibfnamefont {M.}~\bibnamefont {Lucchini}}, \bibinfo {author} {\bibfnamefont {W.}~\bibnamefont {Chung}}, \bibinfo {author} {\bibfnamefont {S.}~\bibnamefont {Eno}}, \bibinfo {author} {\bibfnamefont {Y.}~\bibnamefont {Lai}}, \bibinfo {author} {\bibfnamefont {L.}~\bibnamefont {Lucchini}}, \bibinfo {author} {\bibfnamefont {M.}~\bibnamefont {Nguyen}},\ and\ \bibinfo {author} {\bibfnamefont {C.}~\bibnamefont {Tully}},\ }\bibfield  {title} {\enquote {\bibinfo {title} {New perspectives on segmented crystal calorimeters for future colliders},}\ }\href {https://doi.org/10.1088/1748-0221/15/11/p11005} {\bibfield  {journal} {\bibinfo  {journal} {Journal of Instrumentation}\ }\textbf {\bibinfo {volume} {15}},\ \bibinfo {pages} {P11005--P11005} (\bibinfo {year} {2020})}\BibitemShut {NoStop}%
\bibitem [{\citenamefont {{Chung, Wonyong}}(2025)}]{calor2024}%
  \BibitemOpen
  \bibfield  {author} {\bibinfo {author} {\bibnamefont {{Chung, Wonyong}}},\ }\bibfield  {title} {\enquote {\bibinfo {title} {Differentiable full detector simulation of a projective dual-readout crystal electromagnetic calorimeter with longitudinal segmentation and precision timing},}\ }\href {https://doi.org/10.1051/epjconf/202532000052} {\bibfield  {journal} {\bibinfo  {journal} {EPJ Web Conf.}\ }\textbf {\bibinfo {volume} {320}},\ \bibinfo {pages} {00052} (\bibinfo {year} {2025})}\BibitemShut {NoStop}%
\end{thebibliography}%
\end{document}